# Edge Computing and Dynamic Vision Sensing for Low Delay Access to Visual Medical Information

Ziyang Chen, *Tamanna Shikh-Bahaei, Mohammad Shikh-Bahaei*

*Abstract*—A new method is proposed to decrease the transmission delay of visual and non-visual medical records by using edge computing and Dynamic Vision Sensing (DVS) technologies. The simulation results show that the proposed scheme can decrease the transmission delay by 89.15% to 93.23%. The maximum number of patients who can be served by edge devices is analysed.

## I. INTRODUCTION

Electronic Medical Records (EMR) integrate patients' clinical data and constitute a main source of reference for their care [1]. It is very important for the hospital or GP to access patient's EMR in order to provide timely medical care [2]. An important trend in medical informatics is the adoption of electronic patient record systems that facilitate access to clinical information and work toward preventing the loss or misplacement of information [3, 4].

In the past decades, many telecommunication technologies have been proposed and used to process patients' electronic medical records. With the advanced telecommunication technologies, the patients' medical records are no longer paper-based and can be well organized and shared among the hospitals, which is essential in delivering a good quality of medical care [5].

Cloud computing is one of the important technologies in the healthcare sector. In [6], cloud computing is proposed to collect patients' data. The purpose of the proposed design is to reduce the possibility of typing mistakes in the process of medical data collection and to provide always-on, real-time data collection as opposed to the manual handling of patient information. In [7], cloud computing is proposed to share patient medical records. The aim in [7] is to provide flexible access for patients and different medical professionals. Optimal adaptive and cross layer methods in transmission of data over wireless networks have been studied in the literature [8-11].

Cloud computing has been widely applied in various areas and has proven to be an effective method to collect and process all the data in the cloud. For the terminal users (i.e. mobile users, computer users etc.), different types of terminal devices can access the cloud easily and share data via the cloud [12]. So by applying cloud computing in healthcare area, it's a great convenience for patents, family and medical care workers who want to access EMR immediately. For the service provider (i.e. medical care provider), the cloud is highly scalable. Larger scale services can be expanded easily if more service is demanded [13]. Also, cloud computing can be easily implemented.

The paradigm of cloud computing applied in healthcare is shown in Figure 1. The medical data from various sensors is collected by medical care gateway (i.e. smart phone or other mobile devices). Medical care gateway transfers medical data to cloud. Medical data is processed and analysed in the cloud. After medical care providers and family received requirements from cloud, instructions，suggestions and configurations would send back to patients.

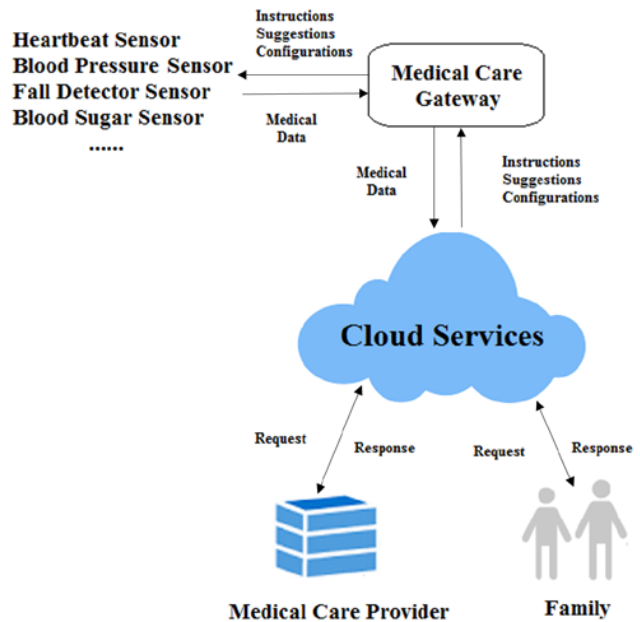

Figure 1. The paradigm of cloud computing

However, there's an increase in the amount of data correlating with the rapidly increasing number of IOT (internet of things) devices. The speed of data transportation is becoming the bottleneck for the Cloud based computing paradigm [14]. For example, wearable sensors and monitoring cameras produce huge amounts of image or video data every second and add a heavy traffic burden to the current networks. Cloud computing, especially, is not efficient enough for those health devices that require very short response times, thus could cause intolerable network latency.

On the other hand, the data which is produced by wearable sensors or other health devices, is usually private and confidential. Transferring and processing patient information through public cloud would pose a breach of patient confidentiality.

Ziyang Chen, Ph.D is with the Informatics Department, King's College London, London, UK (corresponding author to provide e-mail: Ziyang.chen@kcl.ac.uk , telephone number: +44(0)7562845158).
Tamanna Shikh-Bahaei is with Barts and the London, School of Medicine and Dentistry, Queen Mary University of London, London, UK (e-mail: t.shikh-bahaei@smd14.qmul.ac.uk).
Mohammad Shikh-Bahaei, Dr. is with the Informatics Department, King's College London, London, UK (e-mail: m.sbahaei@kcl.ac.uk).

Here we propose edge computing to cache and share medical records. Edge Computing enables mobile subscribers to access IT and cloud computing services at a close proximity within the range of Radio Access Network (RAN) [15]. In the traditional cloud computing, all the information and data are transferred to the cloud network, then users would receive the data through the corresponding base station. Comparing with the traditional cloud computing, edging computing has three main advantages:

- Edge computing could cache and process computing task at the edge of the network without transferring to the cloud network. Edge computing can offload part or all of network traffic from the cloud network to the edge, which would drop the network latency and decrease the bandwidth consumption [16].

- For the issue of privacy, caching patients' health data at the edge is safer. Patients can have their own medical information at a close proximity and control it should the health data be transferred to cloud and service providers.

- Edge computing saves expense and does not need more network infrastructure. For a patient, mobile phone is the edge between body sensors and the cloud. For a smart home, gateway or Wi-Fi is the edge between home sensors and the cloud. [14]

The paradigm of edge computing is shown in Figure 2. We consider the scenario where edge computing offloads all the network from cloud firstly. Raw medical data from various sensors is sent to edge devices. Before being sent to cloud, medical data is cached and processed at the edge. Medical care providers and family would send instructions, suggestions and configurations back after received requirements from cloud. In this scenario, cloud is used to transfer processed data only.

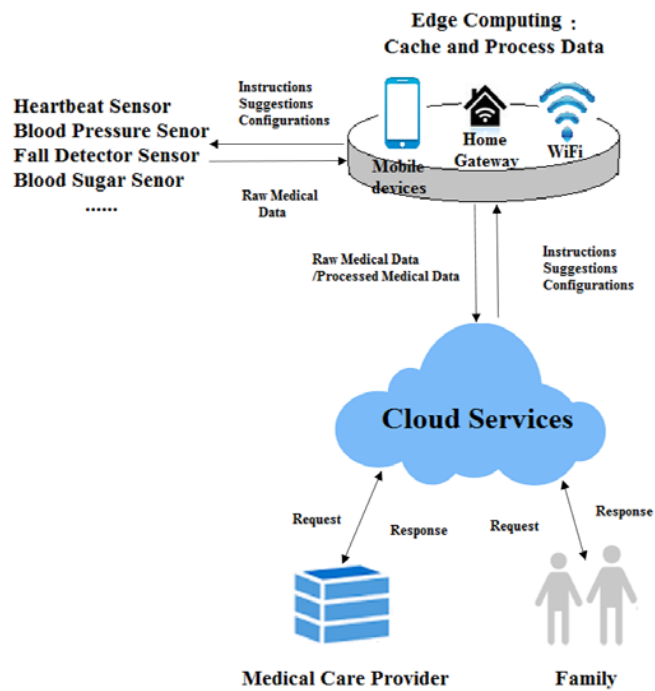

Figure 2. The paradigm of edge computing

For the scenario that edge computing offload part of the network from cloud, part of raw medical data is processed on the edge. The rest is sent to cloud. Medical care providers and family send responses back after receive requirements from cloud. In this scenario, cloud is used to transfer processed data and process raw medical data both. Edge computing can increase the edge responsibility and allows computation and services to be hosted at the edge [17].

Previously Ziyang [18] proposed a new medical files allocation algorithm by using femtocaching to transfer and cache EMRs. The results show a decrease in transmission delay by 44% to 91% in the newly proposed scenario. Compared to femtocaching, edge computing has greater security of data, does not require infrastructure and before the data is sent to the core network it is analysed. Also, in paper [18], EMR is divided into three levels according to the type and size of the files (text/word, digital image and video). However, the size of video (e.g. patient monitoring), accounts for a big part of the EMR (around 70%), which causes massive delay when transferring the EMR.

TABLE I. THE THREE LEVELS OF EMR

| EMR | The medical files |
| --- | --- |
| The first level | Text/word |
| The second level | Image |
| The third level | Video/audio |

In this paper, EMR is still divided into three levels. However, another technology is proposed to decrease the video size by using a dynamic vision sensor camera [19]. This camera not only saves transmission time for the patient, but also saves more space for other patients who need to cache their medical files.

In particular, for the first time, we will be exploiting the recently developed dynamic vision sensor (DVS) technology to achieve smaller medical visual data files in our proposed tele-health system. DVS is different to the conventional video camera; it can detect temporal contrast of brightness and contains an array of asynchronous autonomous self-signaling pixels which respond quickly to relative changes according to the intensity of light [20]. The conventional video cameras are frame based, which means the camera captures a certain number of still images in one second and rapidly plays it back, giving the viewer the illusion of motion [21]. Conventional video cameras have two main disadvantages:

- The first being the limited number of frames which will miss most of the movement if the objects move too fast [22]. For example using a conventional video camera with 30 frames per second to record the punches of a boxer, the camera cannot avoid the problem that all the image sensors have the same timing source. This weakness leads to inadequate data when analysing nuances of the boxer's arm motion. Higher frames and sophisticated processing may improve this case, however, limited bandwidth, power and computing resources again limit this possibility.

- The second disadvantage is the huge storage capacity required due to the large quantity of data produced by

these cameras. The real-time tracking system requires large computational effort and is consequently done on high-performance computer platforms [23]. For example in using a conventional video camera to monitor and record the sleep pattern of a narcolepsy patient at home, only the data relating to the patients' movements are needed. However, huge unvalued and redundant data is included when the subjects are not moving, the furniture and other unchanging parts of the room. This significantly increases processing time and causes a waste of storage space.

In [24], the author uses DVS camera to track the real-time movement of vehicles. The results show it can reduce the computation burden significantly compared to traditional traffic surveillance systems. Compared with the conventional camera, the DVS camera has the fastest response time in recognising fast motion. As mentioned before, the conventional camera is frame-based. However, the DVS camera is event-based. The rate at which frames are captured is entirely dependent upon the rate of light intensity changes, which not only provides more accurate analysis for the fast motions, but also decreases the use of storage capacity, bandwidth and power. A DVS camera typically requires orders of magnitude lower storage capacity than that of a conventional frame-based camera [21]. The working pattern of DVS is shown in the following chart.

TABLE II. THE WORKING PATTERN OF DVS

| Motion of objects | DVS camera working pattern |
|---|---|
| Fast moving | Taking samples with higher rates |
| Slow moving | Taking samples with lower rates |
| No changes | No samples being taken |

In this paper:

- Edge computing is proposed to process and cache EMR. Comparing traditional cloud computing, Edge computing could provide more efficient and safer services. Patient can control their own medical data, made by wearable sensors, on the edge.

- In particular, for the first time, the recently developed dynamic vision sensor (DVS) technology is exploited to achieve smaller medical visual data files in our proposed health system.

- A medical files allocation algorithm is proposed to cache and allocate EMR on the edge. Patient and clinical people can access medical records in a timely manner, and transmission delay is reduced when the medical data is being shared and allocated among the hospital.

- A scenario that patients share their storage capacity of edge devices is considered. More patients could get efficient and effective medical care services.

## II. THE NEW SYSTEM TO ALLOCATE AND CACHE EMR

A new system is proposed to decrease the medical video size and allocate a patient's EMR by using edge computing and DVS camera.

### A. New System for recording and transmitting medical visual and non-visual data

In proposed system we assume that the registered hospital has the complete EMR of the patients. Patients use wearable sensor devices to monitor their health condition and record their location.

The wearable sensor has two main functions. The first one is the detection of emergency situations (e.g. falls). In an emergency situation, an electronic impulse produced by patient's body may be above or below the critical value. The sensor would then send an emergency message to the hospital and clinical professionals. The patients can receive immediate medical services [15]. The caregivers and family of the patient are notified when the emergency situation happens. Also, other notifications would be sent in many cases, for example, if the patient requires assistance in taking their medicine. Clinical personnel can remotely monitor patient's status and be alerted in case a medical decision has to be made[26]. The second main function of the sensor is the recording of the patient's location. Using this approach, a warning about the fall and the location of the subject undergoing monitoring is transmitted to a caregiver or family member via SMS, email and Twitter messages, etc. Also, this approach can record how long the patient stays in this location.

In this monitoring system, we install DVS cameras and wearable sensors in the home area (e.g. tracking the sleeping pattern for the patient who is suffering from a sleep disorder). Also, the DVS camera and wearable sensors are located in their workplace (e.g. for tracking the diet of the diabetic patient). In other places, patients use wearable sensors to record current health situations.

As mentioned before, video records account for most of the space of EMRs. The size of video files, such as those for recording falls or monitoring Alzheimer patients, can be hundreds of Giga bites, especially if a high-resolution camera is deployed. In our daily lives, CIF resolution (352x240) is typically used by mid-level stand-alone DVR recorders when recording real time video [27]. It is also typically used by higher end systems for remote Internet viewing. Most monitoring and surveillance cameras use CIF resolution. The bit rate of a CIF camera is 512kbps. So in the proposed system, we use 352x240 CIF with DVS to record the sleep behavior pattern for a whole night. INILABS, a company specialized in DVS technology, [28] studied sleep behavior pattern by using a 128x128 DVS camera. The DVS only outputs the subjects' movements. A whole night of sleep can be recorded in 100 MB of storage and played back in less than a minute. Activity levels can be automatically extracted and any part of the recording can be viewed at one millisecond resolution. The bit rate for 128x128 DVS camera is suggested to be 256kbps. So the size of the video made by 352x240 CIF camera with DVS would be around 200MB. However, the size of recorded visual data using conventional CIF resolution cameras is around 2.35GB (512x12x3600/8/1024) for a whole night.

We study a scenario with limited edge caching capacity, and for quantitative analysis we assume that video records with conventional CIF resolution camera require a storage capacity of 200GB. So the size of video records made by CIF resolution camera with DVS would be 16.66GB.

There are 6 parts (The registered hospital, home area, work place, family's home, friend's home and other places) in the proposed scenario, as shown in Figure 3. The registered hospital has the entire EMR of the patient and would transfer medical records anytime if other hospitals required it. Edge device A (EA) and edge device B (EB), are located near patient's home and workplace respectively. In the home area and work, DVS cameras and wearable sensors are used to monitor the patient's health status. Medical data is processed on the EA and then transferred to the macro base station. Also, if anything happens, the edge devices would send emergency messages to the nearest hospital. The nearest hospital would check the situation of patients and access the EA to get the health information. In the work place, DVS cameras and wearable sensors are used as well, which has same function as the one in the home area. The EA and EB can process the medical data and cache the most effective medical records on the edge (Medical Records Allocation Algorithm, mentioned in next part). The registered hospital has the complete EMR of the patient and updates EMR accordingly. The data produced by DVS cameras and wearable sensors are stored in edge caches and transmitted to the registered hospital during low data traffic times, e.g. at night. Also edge caches can request the rest medical files from the registered hospital to provide more effective medical care.

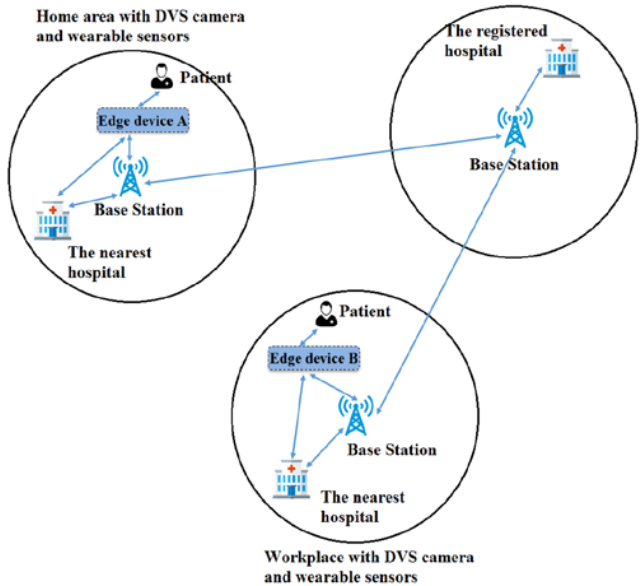

Figure 3.  The proposed system in home area and workplace

The Figure 4, shows the whole of the scenario with 6 parts. Edge computing C (EC), edge computing D (ED) and edge computing E (EE), are respectively located near each family's home, friend's home and other places. In those three parts, only wearable sensors are used to monitor the patient's health status due to a shorter visiting time. Each edge still has the function to allocate, transfer and cache medical records.

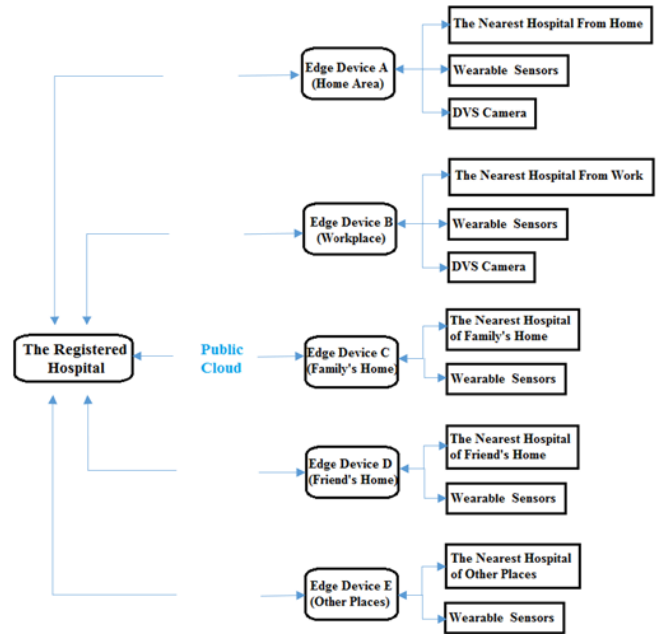

Figure 4.  The proposed system in home area, workplace, family's home, friend's home and other places

Here we consider a typical case where a patient spends, on average, 10 hours at home, 8 hours at work, 3 hours at the family home, 2 hours at their friend's home and 1 hour in other locations. We combine the DVS camera and edge computing technologies to decrease the processing time and delay, and improve the transmission efficiency of medical files. In the next part, a medical files allocation algorithm is proposed to allocate the most important patient medical files to cache in edge devices. Also, we considered a situation in which the patients can share the same edge device. The results show how many more patients the proposed system can serve. Results show a greater number of patients can be served this way.

### B.  Medical Files Allocation Algorithm on the Edge

Due to the limiting storage capacity, each edge device is proposed to select the most important and suitable medical files for the hospital near home, work place and other places [24]. We assume the capacities of edge devices are vary. In system, the storage capacities of EA, EB, EC, ED and EE are 100GB, 500GB, 150GB, 50GB and 10GB, respectively. The ideal situation would be such that physicians can check patient's health condition without needing to request it from their registered hospital. An effective medical files caching algorithm is proposed by deploying the Knapsack model with penalty minimization [18] for optimizing the process of file placement. Each Medical file is correlated to a series of penalty parameters. In proposed system, the staying time of patients in each location is considered as a factor in optimizing the caching process. Also which medical file has the best value for clinical professionals according to the disease condition and the file size is considered.

The optimization problem is to minimize the overall penalty in order to determine the EMR with the lowest penalty and highest priority to be stored in the edge devices [30]. The staying time of patient in different area is considered. The patients who stay longer in one area should be given higher priorities. For example, if the patients spend most of time at home, the edge device of home area should cache more medical files comparing other places where patients stay in with less time. The events of higher priority should be associated with lower penalties [31]. The optimisation problem is modeled by Equation (1):

Minimize:

$$P = \sum_{f \in M} \alpha_f x_f$$

Subject to:

$$x_f = 0 \text{ or } 1 \tag{1}$$

Where

$P$ : Minimum value of penalty.

$M$ : The different medical types from complete EMR.{text, image, DVS video}.

$\alpha_f$ : The penalty coefficient of staying time.

$x_f$ : The decision about allocating medical file f in edge device or not.

The penalty coefficient $\alpha_f$ is presented as in TABLE III.

TABLE III. THE STAYING TIME PENALTY PARAMETERS

| Staying Time (hours) | 1 | 2 | 3 | 4 | … | 21 | 22 | 23 | 24 |
|---|---|---|---|---|---|---|---|---|---|
| Penalty Coefficient; | 24 | 23 | 22 | 21 | … | 4 | 3 | 2 | 1 |

According to the condition of disease, the medical files' value should be considered in allocation algorithm. For example, X-ray images are more important than video files in terms of diagnosing the pathogenic condition of patient who is suffering fractures [32]. In this situation, X-ray images have higher priority and lower penalties than other type of medical files. Therefore, the optimization equation (1) is updated by adding penalty parameter of medical files' value:

Minimize:

$$P = \sum_{f \in M} \alpha_f x_f + \sum_{f \in M} \lambda_f y_f$$

Subject to:

$$x_f, y_f = 0 \text{ or } 1 \tag{2}$$

Where:

$\lambda_f$ : The Penalty coefficient of the medical files' value according to the disease.

$y_f$: The decision about allocating file f in edge devices or not.

In proposed system, the image file is considered as the most important file for clinical people to diagnose the patient's disease. The video file is considered as the least important file. Therefore, the penalty coefficient $\lambda_f$ can be presented as in TABLE IV.

TABLE IV. THE PENALTY COEFFICIENT OF MEDICAL FILES' VALUE

| File Type | Penalty Coefficient $\lambda_f$ |
|---|---|
| Images | 1 |
| Text/word | 2 |
| DVS Video | 3 |

The size of medical files is considered as one of the penalty factors, which is related to transmission time. The more medical files cached in edge devices, the more transmission delay can be decreased. The ideal situation is the edge devices have the whole EMR and no need to request medical files from the registered hospital. DVS camera is used in system to decrease the size of video. However, the capacity storage of each edge device is still limited. Caching more medical files on edge devices could save more transmission time. Therefore, a higher priority and lower penalty should apply when edge devices have more medical files. The optimisation Equation (2) is updated by adding penalty parameters of the transmission delay and limited storage constraint:

Minimize:

$$P_n = \sum_{f \in M} \alpha_f x_f + \sum_{f \in M} \lambda_f y_f + \sum_{f \in M} \beta_f z_f$$

Subject to:

$$\sum_{f \in M} \varphi_{fn} y_{fn} < S_n$$

$$x_f, y_f, z_f = 0 \text{ or } 1 \tag{3}$$

Where:

$\beta_f$: The penalty coefficient of medical files after adding DVS camera.

$z_f$ : The decision about allocating medical file m in edge devices or not by adding DVS.

$\varphi_{fn}$ : The size of the medical records that allocate in edge devices n.

$S_s$ : The maximum storage capacity of each edge device.

As mentioned before, the patient's medical records are defined into text/word, images and DVS video. Therefore, there are seven combinations that edge devices would cache. The penalty coefficient $\beta_f$ can be presented as in TABLE V.

TABLE V. PENALTY PARAMETERS OF TRANSMISSION DELAY

| The Medical Files Cached On Edge | Penalty Coefficient $\beta_{fn}$ |
|---|---|
| Text/word, images, DVS video (106.66G) | 2 |
| Images, DVS video (103.66G) | 4 |
| Text/word, images (90G) | 6 |
| Images (87G) | 8 |
| Text/word, DVS video (19.66G) | 10 |
| DVS Video (16.66G) | 12 |
| Text/word (3G) | 14 |

According to the storage capacity of each edge device, and the penalty coefficients (TABLE III, TABLE IV, TABLE V), the optimization model in Equation (3) is implemented by using LpSolve. The optimized results are shown in TABLE VI.

TABLE VI. THE OPTIMIZED MEDICAL FILES ALLOCATION

| Edge devices | Medical files allocation |
|---|---|
| Edge devices A | Text/words, Image (90G) |
| Edge devices B | Text/word, Images, DVS Video(106.66G) |
| Edge devices C | Text/word, Images, DVS Video(106,66G) |
| Edge devices D | Text/words (3G) |
| Edge devices E | Text/words (3G) |

## III. MODELING AND SIMULATION ANALYSIS

The medical record transmission delay [33] in the proposed system can be presented as follows:

$$D = P_H D_{TH} + P_W D_{TW} + P_{FM} D_{TFM} + P_{FD} D_{TFD} + P_O D_{TO}$$

$$= P_H (\frac{N_{H1}}{R_1} + \frac{N_{H2}}{R_2}) + P_W (\frac{N_{W1}}{R_1} + \frac{N_{W2}}{R_2}) + P_{FM} (\frac{N_{FM1}}{R_1} + \frac{N_{FM2}}{R_2})$$

$$+ P_{FD} (\frac{N_{FD1}}{R_1} + \frac{N_{FD2}}{R_2}) + P_O (\frac{N_{O1}}{R_1} + \frac{N_{O2}}{R_2})$$

$$(4)$$

Where

$D_T$: The transmission delay.

$P_H$, $P_W$, $P_{FM}$, $P_{FD}$, $P_O$ are respectively the possibilities that patient spends time at home, work, family's home, friend's home, and other places.

$D_{TH}$, $D_{TW}$, $D_{TFM}$, $D_{TFD}$, $D_O$ are the transmission delay when medical files are requested from each edge device of home, workplace, family's home, friend's home and other places respectively.

$N_{H1}$, $N_{W1}$, $N_{FM1}$, $N_{FD1}$, $N_{O1}$ are respectively the medical records which are allocated in edge device A, B, C, D and E in bits.

$N_{H2}$, $N_{W2}$, $N_{FM2}$, $N_{FD2}$, $N_{O2}$ are the rest of medical records in bit from registered hospital.

$R_1$: The transmission rate of edge devices.

$R_2$: The transmission rate of macro cellular network.

According to the Poisson distribution [34], the average transmission delay of medical records with caching can be presented as:

$$D = \frac{\sum_K \frac{e^{-\lambda_H} \lambda_H{}^K}{K!}}{K}(\frac{N_{H1}}{R_1} + \frac{N_{H2}}{R_2}) + \frac{\sum_K \frac{e^{-\lambda_W} \lambda_W{}^K}{K!}}{K}(\frac{N_{W1}}{R_1} + \frac{N_{W2}}{R_2})$$

$$+ \frac{\sum_K \frac{e^{-\lambda_{FM}} \lambda_{FM}{}^K}{K!}}{K}(\frac{N_{FM1}}{R_1} + \frac{N_{FM2}}{R_2}) + \frac{\sum_K \frac{e^{-\lambda_{FD}} \lambda_{FD}{}^K}{K!}}{K}(\frac{N_{FD1}}{R_1} + \frac{N_{FD2}}{R_2})$$

$$+ \frac{\sum_K \frac{e^{-\lambda_O} \lambda_O{}^K}{K!}}{K}(\frac{N_{O1}}{R_1} + \frac{N_{O2}}{R_2})$$

$$(5)$$

Where

$\lambda_H$: The length of staying time by possibility in home area. $K$: The number of occurrences.

$\lambda_W$: The length of staying time by possibility in workplace.

$\lambda_{FM}$: The length of staying time by possibility in family's home.

$\lambda_{FD}$: The length of staying time by possibility in friends' home.

$\lambda_O$: The length of staying time in other places.

The other simulation parameters of the studied scenario:

- The proposed scenario is covered by a single 3GPP LTE R8 cell. The edge devices use a simplified 802.11n protocol.

- The edge device is implemented near patients within 100 meters.

- The values of $\lambda_H$, $\lambda_W$, $\lambda_{FM}$, $\lambda_{FD}$, $\lambda_O$ are 0.4167, 0.3333, , 0.125, 0.0833, 0.0417 respectively.

- The values of $N_{H1}$, $N_{W1}$, $N_{FM1}$, $N_{FD1}$, $N_{O1}$ are shown in TABLE VI, which are 90G, 106.66G, 106,66G, 3G, 3G respectively.

- The values of $N_{H2}$, $N_{W2}$, $N_{FM2}$, $N_{FD2}$, $N_{O2}$ are the rest of EMR, which are 16.66G, 0G, 0G, 103.66,G 103.66G respectively.

The simulation results are analyzed by MATLAB and presented in the following section.

## IV. SIMULATION RESULTS

The simulation results of transmission delay are compared with our previous research [18]. In our previous research, only femtocaching was used in system. We combined edge caching with DVS camera technology in this work to decrease the size of medical and improve transmission delay. We still considered two scenario: the best situation and the worst situation. The best situation corresponds to the case that the edge caches have the enough medical files. The nearest hospital can provide effective medical services and does not need to access the whole of EMR which cached in registered hospital. The transmission delay by using previous system (femtocaching only) is 16.59 minutes. However the transmission delay by applying edge computing and DVS camera technology is 9.872 minutes, which is a 40.5% improvement.

The worst situation corresponds to the case that the medical files from edge caches are not enough for the nearest hospital. The nearest hospital in each location need more complete medical data from the registered hospital to check the patients' health condition. Comparing with the best situation, the transmission delay would increase. The transmission delay by using previous system (femtocaching only) is 139.652 minutes. However the transmission delay by applying edge computing and DVS camera technology is 26.855 minutes, which is an 80.77% improvement. The results are shown in Figure 4.

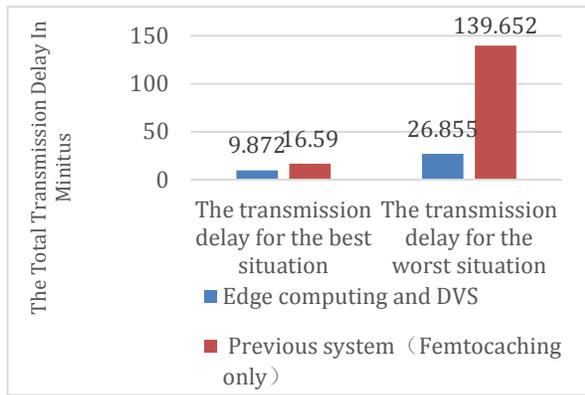

Figure 4. The figures to show how many delays we can save compared to the previous system

We also compared the transmission delay of "Without edge computing and DVS" and "With edge computing and DVS". The results are shown in Figure 5. For the best situation, the transmission delay without using the proposed system is 145.73 minutes. However the transmission delay with DVS and edge computing is 9.872 minutes, which is a 93.23% improvement. For the worst situation, the transmission delay without using the proposed system is 247.467 minutes. However the transmission delay with DVS and edge computing is 26.855 minutes, which is a 89.15% improvement. So the proposed system can improve the transmission delay by 89.15% to 93.23%.

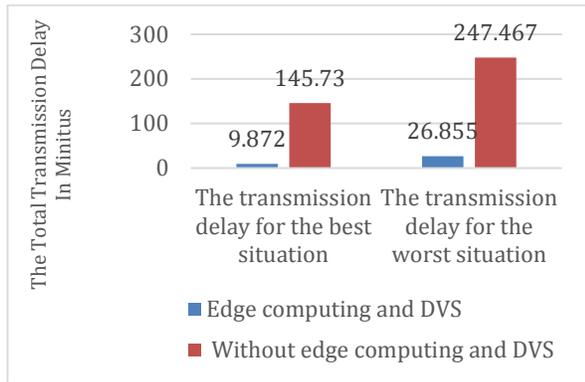

Figure 5. The figures to show how many delays we can save compared to the system without edge computing and DVS

In the proposed system, each patient (which call "Host") has their own edge device. We have considered the scenarios in which patients who live nearby (we call "Guest") can share the whole, part and none of the storage capacity of edge devices. The maximum number of patients who can be served is analyzed. In order to find the maximum number of patients that can be served, we have to guarantee that the "Host" has extra space in their edge device after his/her complete EMR is cached. Also, we considered the situation in which the "Guest" only caches the smallest size of medical files (text/word). In the proposed scenario, the storage capacity of EA, ED and EE is 100G, 50G and 10G respectively. The three edge devices don't have enough space to cache a complete EMR, thus can't be shared with other patients. The storage capacity of EB and EC is 500G and 150G. The number of patients that can be served when edge devices are shared is 132 ((500-106.66)/3+1) and 15 ((150-106.66)/3+1). So in the proposed system, if the edge devices can be shared, the maximum number of patients can be served is 147. Also, the maximum number of patients that can be served improved while the storage capacity of shared edge devices increased. The results are shown in Figure 6. The blue line shows that only host patient could be serviced if edge devices cannot be shared with others. However, the number of patients would increase significantly when edge devices can be shared.

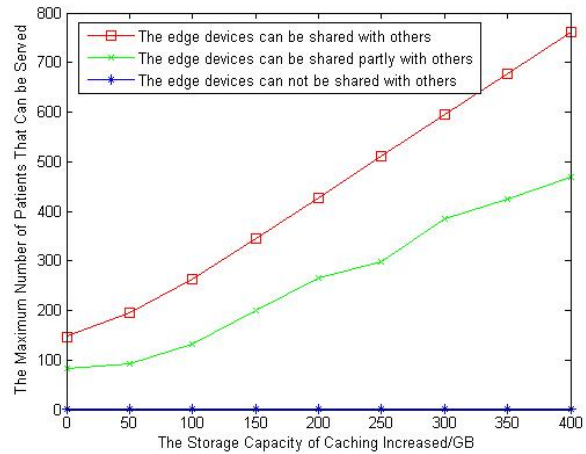

Figure 6. The number of patients can be served when storage capacity of shared edge devices increased

## V. CONCLUSION

In this research, we introduced a new network element on the edge to cache, allocate and transfer patients' medical records. DVS technology is proposed to decrease the size of medical videos to drop more transmission delay. We also compared the proposed system with our previous work. The simulation results show the transmission delay has a improvement by 40.5% to 80.77%. We also compared the situation without using edge computing and DVS. The simulation results show that the proposed system can drop the transmission delay by 89.15% to 93.23%. Also we analysed the situation that edge devices can be shared with other patients, the result shows the maximum number that can be served is 147. The maximum number is increased when the storage capacity of edge devices is improved. In this paper, there are still some issues that can be improved. For example, patients who share their edge devices if edge devices have extra capacity storage. However, there is the issue of security as well as the issue of sacrificing one's battery life and having to deal with a longer processing time if one is to share the edge device. A clear reward algorithm is needed.